\documentclass[trackchanges]{aastex701}
\usepackage[utf8]{inputenc}
\usepackage{comment}

\definecolor{meridithgreen}{RGB}{0, 150, 0}

\begin{document}

\title{Empirical Globular Cluster Ridgeline Construction on \textit{Gaia} Data} 

\author[orcid=0009-0008-5224-4820,gname=Anett, sname='Simon-Zsók]{Anett Simon-Zsók} 
\affiliation{HUN-REN CSFK, Konkoly Observatory, MTA Centre of Excellence, Konkoly Thege Mikl\'os \'ut 15-17, Budapest, 1121 Hungary}
\affiliation{ELTE E\"otv\"os Lor\'and University, Institute of Physics and Astronomy, P\'azm\'any P\'eter s\'et\'any 1, Budapest, Hungary}
\email[show]{simonzsok.anett@csfk.org}

\author[orcid=0000-0002-8159-1599,sname='László Molnár']{László Molnár}
\affiliation{HUN-REN CSFK, Konkoly Observatory, MTA Centre of Excellence, Konkoly Thege Mikl\'os \'ut 15-17, Budapest, 1121 Hungary}
\affiliation{ELTE E\"otv\"os Lor\'and University, Institute of Physics and Astronomy, P\'azm\'any P\'eter s\'et\'any 1, Budapest, Hungary}\email[show]{molnar.laszlo@csfk.org}  

\author[0000-0002-1663-0707]{Csilla Kalup}
\affiliation{HUN-REN CSFK, Konkoly Observatory, MTA Centre of Excellence, Konkoly Thege Mikl\'os \'ut 15-17, Budapest, 1121 Hungary}
\affiliation{ELTE E\"otv\"os Lor\'and University, Institute of Physics and Astronomy, P\'azm\'any P\'eter s\'et\'any 1, Budapest, Hungary}
\email{kalup.csilla@csfk.org}

\author[0000-0002-8717-127X]{Meridith Joyce}
 \affiliation{University of Wyoming, 1000 E University Ave, Laramie, WY USA}
\affiliation{HUN-REN CSFK, Konkoly Observatory, MTA Centre of Excellence, Konkoly Thege Mikl\'os \'ut 15-17, Budapest, 1121 Hungary}
\email{mjoyce8@uwyo.edu}

\begin{abstract}

We present a new method and a corresponding code to compress the color magnitude diagram of a globular cluster into a representative curve, called a ridgeline, that can be more readily compared to isochrone models, among other applications. This compression method preserves the physical properties of the cluster, including the morphology of the CMD.
\end{abstract}


\section{Introduction} 

Globular clusters (GCs) are some of the oldest objects in the known Universe. To first order, stars in a GC can be assumed to have formed at the same time and from the same molecular cloud, making them powerful diagnostics of stellar evolution and absolute age anchors. However, determining the ages of the clusters as an ensemble or through individual stars is still an unresolved problem \citep{Tayar-Joyce-2025}. Isochrone fitting remains the most widely-used method to determine the age, metallicity, or reddening maps of GCs using their color-magnitude diagram (CMD). 

The literature contains many methods for isochrone fitting to CMDs.
One way is to create synthetic CMDs based on isochrones and initial mass functions and compare those to the observed CMDs. \citet{yingM92} use Voronoi tessellation for comparison, while \citet{Souza-2020} employe Bayesian statistics. 

A more classical approach is to compress the GC CMD data into a ridgeline. While this reduces the information content, it is much less computationally expensive than the methods mentioned above. However, this is a non-trivial problem because the slope of the CMD varies greatly and non-monotonically, especially from the main sequence (MS) turnoff point to the base of the Red Giant Branch (RGB), which are very important diagnostic features. 
As such, simple binning along either the color or brightness axis will cause issues with insufficient sampling of these regions. While algorithmic solutions exist \citep[see, e.g.,][]{Alonso_Garc_a_2011}, ridgelines have also been defined by visually connecting defined anchor points, or from hand-drawn fiducial lines \citep[e.g.,][]{Stetson-2005,Zorotovic-2009}. 

\begin{figure*}[ht!]
  \centering
  \includegraphics[width=\linewidth, trim=0 0 0 0, clip]{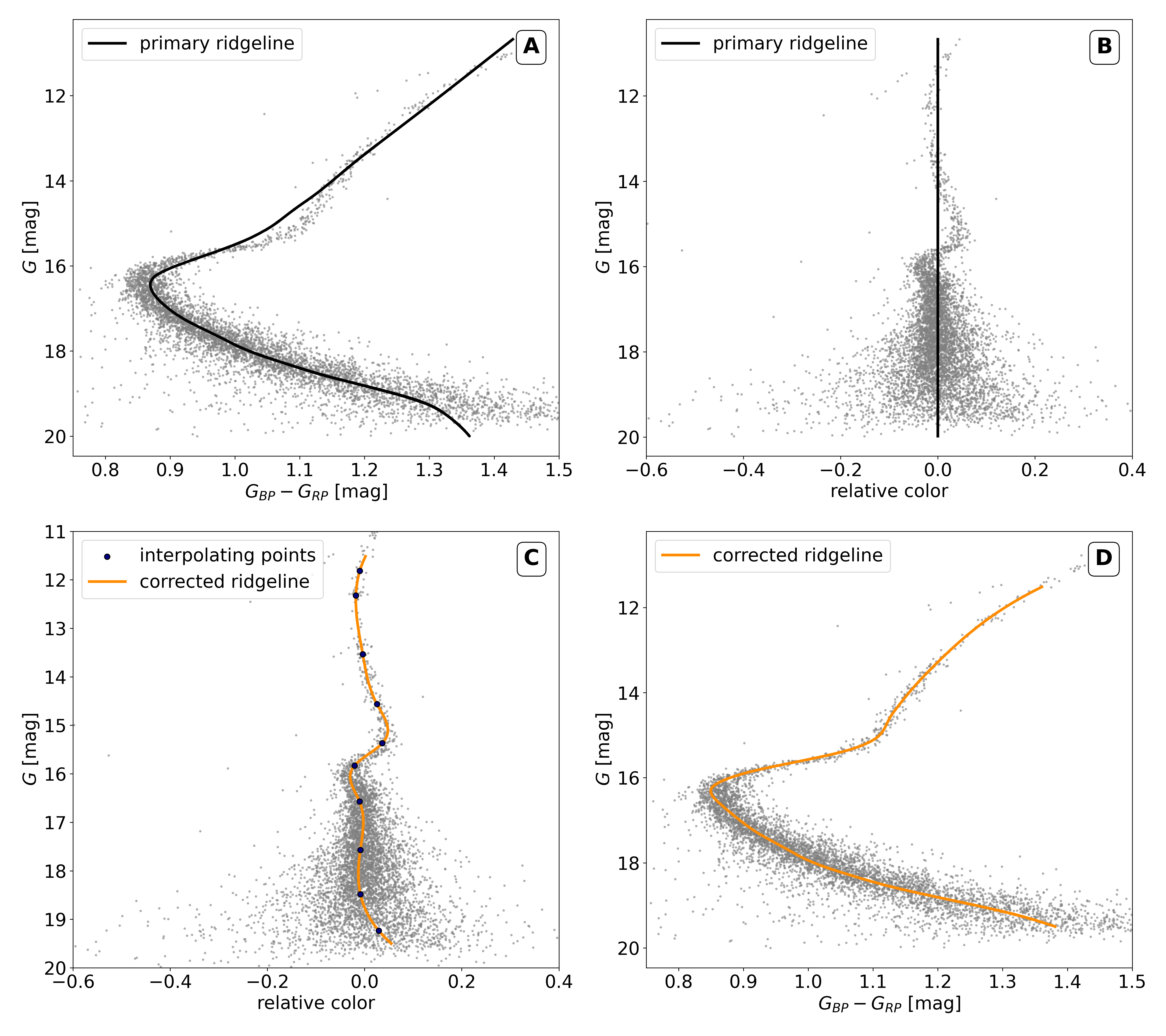}
  \caption{The ridgeline determination workflow in case of globular cluster NGC~6397 as an example. \textbf{A:} Primary ridgeline after applying the LOWESS algorithm along the Y axis. \textbf{B:} The CMD and the primary ridgeline in the transformed coordinate system. \textbf{C:} The the X coordinate of the blue points are maximum value of the KDE in a bin. The Y coordinates are the median values for the bins. The orange line is the cubic interpolation along the blue points. \textbf{D:} The final ridgeline after the rotating back to the original coordinate system.}
  \label{fig:general}
\end{figure*}

\section{Gaia Data}

The algorithms mentioned above are not available publicly. In this work we present a relatively simple algorithm and \texttt{python} code to calculate ridgelines. We have chosen to use \textit{Gaia} photometric data because \textit{Gaia} covered the entire sky and thus all the brighter GCs homogeneously and includes stellar proper motions as well. The necessary parameters of the stars (in an arbitrary rectangular projection of the sky) can be easily downloaded and post-processed with the help of the \texttt{astroquery.gaia} module \citep{Astropy}.

After identifying the cluster members based on proper motion data, reductions were applied to correct for instrumental effects. Using the color excess parameter ($C^*$) defined by \citet{Riello_Gaia}, we applied a 1$\sigma$ cut in $C^*$ (as a function of $G$ brightness) to exclude stars with potentially problematic photometry. Due to the overestimated mean flux for faint red sources, we also applied the $G_{BP}<20.3$~mag local flux threshold to the data \citep{Riello_Gaia}. 

\section{Method}

We work in two coordinate systems. The first is the system of the Gaia CMD: $G_{BP}-G_{RP}$ color on the $X$ axis, and $G$ on the $Y$ axis. 
The second is the reference frame of the ``straightened'' primary ridgeline. Figure~\ref{fig:general} presents a ridgeline determination workflow, applied here to the cluster NGC6397.

The algorithm begins by determining the primary ridgeline. We used the LOWESS (Locally Weighted Scatterplot Smoothing, \cite{Lowess}) method, applied along the $Y$ (brightness) axis of the stars. This produces the line in panel A of Figure \ref{fig:general}. We then ``unfold'' the CMD by rotating the data into a frame where the ridgeline sets the Y axis (Panel B). The transformation is one-to-one between the two coordinate systems: we assign a point along the ridgeline to each star at the same brightness coordinate, and then calculate the instantaneous derivative of the ridgeline at that point. It can be seen that around the turnoff point and on the lower part of the RGB, the line does not capture the morphology of the GC data well. We therefore perform a second fitting. We separate the data into 10 bins along the $Y$ axis. As the distribution of stars along the color ($X$) axis (across the CMD) changes with evolutionary stage, we calculate KDE (Kernel Density Estimate) values 
along the $X$ axis of each bin and determine the color coordinates as the positions of the maxima of the KDEs. For the $Y$ (brightness) coordinates of the bins, along the CMD, we calculate the median values of the $G$ brightness per bin. 
The maxima of the KDEs become the $X$ coordinates and the median values become the $Y$ coordinates of the 10 well-defined bin points. We connect these points using a cubic spline (Panel C). 

In the final step, we transform the cubic spline back into the original coordinate system, applying in reverse the inverses of the transformations we applied between Panels A and B. The interpolated cubic spline thus becomes the final ridgeline, which now includes the CMD features captured both by the initial fit (Panel A) and by the cubic spline (Panel C). This way, the final ridgeline now closely follows the turnoff point and the RGB (Panel D).  

This method works well on single-population globular clusters. It is worth mentioning that if there is a crowded horizontal branch, the stars need to be removed manually or by setting the parameters accordingly.  

\section{Conclusion}
This method provides a robust approach for determining the ridgeline of a CMD of a globular cluster. A ridgeline allows us to perform KS tests with isochrones to quantitatively determine best fit (rather than visually evaluating the fit). This technique can also be used to map the local relative reddening in front of the cluster based on the color offset of stars from the line \citep{Alonso_Garc_a_2011}. 
%
Although this demonstration uses Gaia data, it can be readily adapted to data in any color-magnitude system (e.g., Rubin Observatory).

\begin{acknowledgments}
The code is publicly available on Github: \url{https://github.com/anettszs/GCRidgeline}\\
This research was supported by the ‘SeismoLab’ KKP-137523 grant and the TKP2021-NKTA-64 excellence grant of the Hungarian Research, Development and Innovation Office (NKFIH). A.\,S-Zs.\ acknowledges the undergraduate research assistant program of Konkoly Observatory.

\end{acknowledgments}

%
\facilities{Gaia \citep{Gaia2016}}

\software{astropy \citep{Astropy}, astroquery \citep{astroquery-2019}}

\bibliography{main}{}

\begin{thebibliography}{}
\expandafter\ifx\csname natexlab\endcsname\relax\def\natexlab#1{#1}\fi
\providecommand{\url}[1]{\href{#1}{#1}}
\providecommand{\dodoi}[1]{doi:~\href{http://doi.org/#1}{\nolinkurl{#1}}}
\providecommand{\doeprint}[1]{\href{http://ascl.net/#1}{\nolinkurl{http://ascl.net/#1}}}
\providecommand{\doarXiv}[1]{\href{https://arxiv.org/abs/#1}{\nolinkurl{https://arxiv.org/abs/#1}}}

\bibitem[{J. Alonso-García {et~al.}(2011)Alonso-García, Mateo, Sen, Banerjee, \& von Braun}]{Alonso_Garc_a_2011}
Alonso-García, J., Mateo, M., Sen, B., Banerjee, M., \& von Braun, K. 2011, \bibinfo{title}{MAPPING DIFFERENTIAL REDDENING IN THE INNER GALACTIC GLOBULAR CLUSTER SYSTEM,} AJ, 141, 146

\bibitem[{ {Astropy Collaboration} {et~al.}(2013){Astropy Collaboration}, {Robitaille}, {Tollerud}, {Greenfield}, {Droettboom}, {Bray}, {Aldcroft}, {Davis}, {Ginsburg}, {Price-Whelan}, {Kerzendorf}, {Conley}, {Crighton}, {Barbary}, {Muna}, {Ferguson}, {Grollier}, {Parikh}, {Nair}, {Unther}, {Deil}, {Woillez}, {Conseil}, {Kramer}, {Turner}, {Singer}, {Fox}, {Weaver}, {Zabalza}, {Edwards}, {Azalee Bostroem}, {Burke}, {Casey}, {Crawford}, {Dencheva}, {Ely}, {Jenness}, {Labrie}, {Lim}, {Pierfederici}, {Pontzen}, {Ptak}, {Refsdal}, {Servillat}, \& {Streicher}}]{Astropy}
{Astropy Collaboration}, {Robitaille}, T.~P., {Tollerud}, E.~J., {et~al.} 2013, \bibinfo{title}{{Astropy: A community Python package for astronomy},} \aap, 558, A33, \dodoi{10.1051/0004-6361/201322068}

\bibitem[{A. Derkacheva {et~al.}(2020)Derkacheva, Mouginot, Millan, Maier, \& Gillet-Chaulet}]{Lowess}
Derkacheva, A., Mouginot, J., Millan, R., Maier, N., \& Gillet-Chaulet, F. 2020, \bibinfo{title}{Data Reduction Using Statistical and Regression Approaches for Ice Velocity Derived by Landsat-8, Sentinel-1 and Sentinel-2,} Remote Sensing, 12

\bibitem[{ {Gaia Collaboration} {et~al.}(2016){Gaia Collaboration}, {Prusti}, {de Bruijne}, {Brown}, {Vallenari}, {Babusiaux}, {Bailer-Jones}, {Bastian}, {Biermann}, {Evans}, {Eyer}, {Jansen}, {Jordi}, {Klioner}, {Lammers}, {Lindegren}, {Luri}, {Mignard}, {Milligan}, {Panem}, {Poinsignon}, {Pourbaix}, {Randich}, {Sarri}, {Sartoretti}, {Siddiqui}, {Soubiran}, {Valette}, {van Leeuwen}, {Walton}, {Aerts}, {Arenou}, {Cropper}, {Drimmel}, {H{\o}g}, {Katz}, {Lattanzi}, {O'Mullane}, {Grebel}, {Holland}, {Huc}, {Passot}, {Bramante}, {Cacciari}, {Casta{\~n}eda}, {Chaoul}, {Cheek}, {De Angeli}, {Fabricius}, {Guerra}, {Hern{\'a}ndez}, {Jean-Antoine-Piccolo}, {Masana}, {Messineo}, {Mowlavi}, {Nienartowicz}, {Ord{\'o}{\~n}ez-Blanco}, {Panuzzo}, {Portell}, {Richards}, {Riello}, {Seabroke}, {Tanga}, {Th{\'e}venin}, {Torra}, {Els}, {Gracia-Abril}, {Comoretto}, {Garcia-Reinaldos}, {Lock}, \& {Mercier}}]{Gaia2016}
{Gaia Collaboration}, {Prusti}, T., {de Bruijne}, J.~H.~J., {et~al.} 2016, \bibinfo{title}{{The Gaia mission},} \aap, 595, A1, \dodoi{10.1051/0004-6361/201629272}

\bibitem[{A. {Ginsburg} {et~al.}(2019){Ginsburg}, {Sip{\H{o}}cz}, {Brasseur}, {Cowperthwaite}, {Craig}, {Deil}, {Guillochon}, {Guzman}, {Liedtke}, {Lian Lim}, {Lockhart}, {Mommert}, {Morris}, {Norman}, {Parikh}, {Persson}, {Robitaille}, {Segovia}, {Singer}, {Tollerud}, {de Val-Borro}, {Valtchanov}, {Woillez}, {Astroquery Collaboration}, \& {a subset of astropy Collaboration}}]{astroquery-2019}
{Ginsburg}, A., {Sip{\H{o}}cz}, B.~M., {Brasseur}, C.~E., {et~al.} 2019, \bibinfo{title}{{astroquery: An Astronomical Web-querying Package in Python},} \aj, 157, 98, \dodoi{10.3847/1538-3881/aafc33}

\bibitem[{M. {Riello} {et~al.}(2021){Riello}, {De Angeli}, {Evans}, {Montegriffo}, {Carrasco}, {Busso}, {Palaversa}, {Burgess}, {Diener}, {Davidson}, {Rowell}, {Fabricius}, {Jordi}, {Bellazzini}, {Pancino}, {Harrison}, {Cacciari}, {van Leeuwen}, {Hambly}, {Hodgkin}, {Osborne}, {Altavilla}, {Barstow}, {Brown}, {Castellani}, {Cowell}, {De Luise}, {Gilmore}, {Giuffrida}, {Hidalgo}, {Holland}, {Marinoni}, {Pagani}, {Piersimoni}, {Pulone}, {Ragaini}, {Rainer}, {Richards}, {Sanna}, {Walton}, {Weiler}, \& {Yoldas}}]{Riello_Gaia}
{Riello}, M., {De Angeli}, F., {Evans}, D.~W., {et~al.} 2021, \bibinfo{title}{{Gaia Early Data Release 3. Photometric content and validation},} \aap, 649, A3, \dodoi{10.1051/0004-6361/202039587}

\bibitem[{S.~O. {Souza} {et~al.}(2020){Souza}, {Kerber}, {Barbuy}, {P{\'e}rez-Villegas}, {Oliveira}, \& {Nardiello}}]{Souza-2020}
{Souza}, S.~O., {Kerber}, L.~O., {Barbuy}, B., {et~al.} 2020, \bibinfo{title}{{Self-consistent Analysis of Stellar Clusters: An Application to HST Data of the Halo Globular Cluster NGC 6752},} \apj, 890, 38, \dodoi{10.3847/1538-4357/ab6a0f}

\bibitem[{P.~B. {Stetson} {et~al.}(2005){Stetson}, {Catelan}, \& {Smith}}]{Stetson-2005}
{Stetson}, P.~B., {Catelan}, M., \& {Smith}, H.~A. 2005, \bibinfo{title}{{Homogeneous Photometry. V. The Globular Cluster NGC 4147},} \pasp, 117, 1325, \dodoi{10.1086/497302}

\bibitem[{J. {Tayar} \& M. {Joyce}(2025){Tayar} \& {Joyce}}]{Tayar-Joyce-2025}
{Tayar}, J., \& {Joyce}, M. 2025, \bibinfo{title}{{Star-crossed Clusters: Asteroseismic Ages for Individual Stars Are in Tension with the Ages of Their Host Clusters},} \apjl, 984, L56, \dodoi{10.3847/2041-8213/adcd6f}

\bibitem[{J.~M. {Ying} {et~al.}(2023){Ying}, {Chaboyer}, {Boudreaux}, {Slaughter}, {Boylan-Kolchin}, \& {Weisz}}]{yingM92}
{Ying}, J.~M., {Chaboyer}, B., {Boudreaux}, E.~M., {et~al.} 2023, \bibinfo{title}{{The Absolute Age of M92},} \aj, 166, 18, \dodoi{10.3847/1538-3881/acd9b1}

\bibitem[{M. {Zorotovic} {et~al.}(2009){Zorotovic}, {Catelan}, {Zoccali}, {Pritzl}, {Smith}, {Stephens}, {Contreras}, \& {Escobar}}]{Zorotovic-2009}
{Zorotovic}, M., {Catelan}, M., {Zoccali}, M., {et~al.} 2009, \bibinfo{title}{{The Globular Cluster NGC 5286. I. A New CCD BV Color-Magnitude Diagram},} \aj, 137, 257, \dodoi{10.1088/0004-6256/137/1/257}

\end{thebibliography}
\bibliographystyle{aasjournalv7}

\end{document}